\documentclass[preprint2]{emulateapj}			

\usepackage{graphicx, natbib, txfonts,color}
\usepackage[normalem]{ulem}

\newcommand{\HeI}{He~{\sc{i}}}
\newcommand{\OIV}{O~{\sc{iv}}}
\def\CaII{Ca~{\sc{ii}}}      
\def\Halpha{\mbox{H\hspace{0.1ex}$\alpha$}}
\def\kms{\hbox{km$\;$s$^{-1}$}}
\def\mss{\hbox{m$\;$s$^{-2}$}}
\def\gcm{\hbox{g$\;$cm$^{-3}$}}

\begin{document}

\title{Flocculent flows in the chromospheric canopy of a sunspot}
\shorttitle{Flocculent flows in the chromosphere}
\shortauthors{G. Vissers and L. Rouppe van der Voort}

\author{Gregal Vissers}
\author{	Luc Rouppe van der Voort}
\email{g.j.m.vissers@astro.uio.no}

\affil{Institute of
  Theoretical Astrophysics, University of Oslo, P.O. Box 1029
  Blindern, N-0315 Oslo, Norway\email{g.j.m.vissers@astro.uio.no}}

\begin{abstract}
High-quality imaging spectroscopy in the \Halpha\ line, obtained with the CRisp Imaging SpectroPolarimeter (CRISP) at the Swedish 1-m Solar Telescope (SST) at La Palma and covering a small sunspot and its surroundings, are studied.
They exhibit ubiquitous flows both along fibrils making up the chromospheric canopy away from the spot and in the superpenumbra. We term these flows ``flocculent'' to describe their intermittent character, that is morphologically reminiscent of coronal rain.
The flocculent flows are investigated further in order to determine their dynamic and morphological properties.
For the measurement of their characteristic velocities, accelerations and sizes, we employ a new versatile analysis tool, the CRisp SPectral EXplorer (CRISPEX), which we describe in detail.
Absolute velocities on the order of 7.2--82.4\,\kms\ are found, with an average value of 36.5$\pm$5.9\,\kms\ and slightly higher typical velocities for features moving towards the sunspot than away.
These velocities are much higher than those determined from the shift of the line core, which shows patches around the sunspot with velocity enhancements of up to 10--15\,\kms\ (both red- and blueshifted).
Accelerations are determined for a subsample of features, that show clear accelerating or decelerating behavior, yielding an average of 270$\pm$63\,\mss\ and 149$\pm$63\,\mss\ for accelerating and decelerating features, respectively.
Typical flocculent features measure 627$\pm$44\,km in length and 304$\pm$30\,km in width.
On average 68 features are detected per minute, with an average lifetime of 67.7$\pm$8.8\,s.
The dynamics and phenomenology of the flocculent flows suggest they may be driven by a siphon flow, where the flocculence could arise from a density perturbation close to one of the footpoints or along the loop structure.
\end{abstract}


\section{Introduction}
The advent of imaging spectroscopy at high-resolution solar telescopes using adaptive optics and image post-processing allows the study of solar phenomena at unprecedented spatial, temporal and spectral resolution.
In this paper we analyze a high-quality set sampling the chromosphere above a small sunspot and its surroundings in the \Halpha\ line.
The chromospheric canopy exhibits ubiquitous fast-moving small-scale features constituting flows both towards and away from the sunspot in the superpenumbra, as well as in fibrils anchored in plage only.
We call these features and flows ``flocculent''\footnote{They are not the ``flocculi'' reported in the older solar literature, in reference to unresolved \CaII\ H \& K network
\citep[e.g.][]{1974soch.book.....B}.	
} to characterize their intrinsic intermittency.

Similar blob-like phenomena have recently been observed by
\citet{2006ApJ...648L..67V},	
\citet{2008A&A...486..577S},	
\citet{2010MmSAI..81..693W},	
and 
\citet{2011_Lin_etal},	
and have been interpreted as a manifestation of propagating waves rather than actual mass motion.
With spatial dimensions of 0.5--1\,arcsec or less, the features observed are seen to propagate along the fibrillar canopy structure observed in \Halpha\ and \CaII~8542\,\AA, in some cases with additional periodic lateral displacement, at velocities close to (and in most cases well in excess of) the chromospheric sound speed.

The flocculent flows above the sunspot and in its superpenumbra that are reported on here are reminiscent of coronal rain as well as the inverse Evershed effect, and may constitute mass motions rather than waves.
We therefore complete this introduction with brief reviews of the coronal rain phenomenon and the (inverse) Evershed effect.

\paragraph{Coronal rain}
In coronal loops both (transient) siphon flows \citep[e.g.,][]{
1998SoPh..182...73K, 	
2006A&A...452.1075D, 	
2009ApJ...704..883T}		
and ``blobs'' of cooled down plasma (so-called \emph{coronal rain}) 
precipitating along the legs of coronal loops \citep[e.g.,][]{
2001SoPh..198..325S, 	
2005A&A...436.1067M,  	
2005A&A...443..319D,		
2007A&A...475L..25O}  	
have been observed. 
Typical velocities for coronal rain range between a few tens to a few hundred kilometers per second
\citep{2001SoPh..198..325S, 
2005A&A...443..319D, 
2007A&A...475L..25O}, 
which was corroborated through simulations by
\citet{2005A&A...436.1067M}. 
Recent studies by
\citet{2012ApJ...745..152A}	
and
\citet{2012_Antolin_etal}	
have further expanded the statistics for both off-limb and on-disk cases, respectively, confirming earlier dynamics results.
Catastrophic cooling is generally accepted as explanation of the coronal rain phenomenon.
In this process, as advanced by 
\cite{2001SoPh..198..325S}, 
thermal non-equilibrium leads to condensation of plasma at the loop apex followed by a runaway cooling process, with subsequent downflow of one or more blobs along one or both loop legs.
The siphon flow velocities in coronal loops are similar to the velocities found for coronal rain
\citep[cf.][]{1998SoPh..182...73K, 
2006A&A...452.1075D}, 
but do not result in blobs falling down towards the photosphere.

\paragraph{(Inverse) Evershed effect}
The inverse Evershed effect is a chromospheric inflow of plasma along superpenumbral fibrils towards the sunspot umbra, while the outflow observed in the photospheric penumbra is referred to as the (normal) Evershed effect. 
The latter was first discovered by 
\citet{1909MNRAS..69..454E}	
and has been extensively studied in the years since, leading to relatively well established values for the outflow velocity of the order of a few kilometers per second near the umbra-penumbra boundary increasing up to about 10\,\kms\ at the outer penumbral edge. 
Though several models have been proposed in the past, thermal convective motions are nowadays most widely accepted as the driving mechanism for the Evershed effect
\citep[cf.][]{%
2006A&A...447..343S,	
2007ApJ...669.1390H,	
2009ApJ...691..640R,	
2010mcia.conf..243N, 
2011Sci...333..316S}. 

The existence of the inverse Evershed effect was already suggested by
\citet{1909Obs....32..291E}	
following his discovery of the photospheric outflow and more extensively studied by
\citet{1913ApJ....37..322S},	
who confirmed the earlier findings of Evershed that in the chromosphere the flow reverses sign with respect to the photosphere.
Siphon flows are most often proposed as the mechanism behind the inverse Evershed effect
\citep[cf.][]{
1969SoPh....9...88H,	
1975SoPh...43...91M,	
1988A&A...201..339A,	
1990A&A...233..207D,	
1992A&A...259..307B,	
1993SoPh..145..257K,	
1997Natur.390..485M},	
with only few alternatives, such as gravitationally driven in- and downflow 
\citep[][]{1962AuJPh..15..327B}	
or some mechanism analogous to the moving flux tube model
\citep[][]{2008A&A...491L...5T}.	
A wide range in velocities is found, however.
\citet{1913ApJ....37..322S}	
found the inverse Evershed effect to propagate with about 3\,\kms\ on average and similar velocities were obtained by
\citet{1978SoPh...57...65B},	
while somewhat higher velocities were found by
\citet{1969SoPh....9...88H},	
\citet{1990A&A...233..207D}	
and
\citet{1988A&A...201..339A}.	
However, measurements of single absorbing elements in the \Halpha\ spectra analyzed by 
\citet{1969SoPh....9...88H}	
resulted in velocities of up to 50\,\kms\ and even higher velocities were obtained in filtergram studies by
\citet{1962AuJPh..15..327B}	
and
\citet{1975SoPh...43...91M}.	
This difference between spectroscopic and filtergram velocities was already pointed out before by 
\citet{1988A&A...201..339A}.	
The inverse Evershed effect has also been observed in the transition region, albeit at typically higher Doppler velocities on the order of a few tens of \kms\ obtained from spectral analyses
\citep{1988A&A...201..339A,	
1990A&A...233..207D,	
1993SoPh..145..257K,	
2008A&A...491L...5T},	
and it has not been as extensively studied as in the chromosphere. 

In summary, the basic picture that arises from previous studies of the (inverse)		
Evershed effect is one of relatively slow penumbral outflow in the photosphere, a faster inflow in the superpenumbra at chromospheric levels, and an even faster inflow in the transition region. 
\newline\newline
In this paper we describe, measure, and interpret the flocculent flows in our data.
The measurements on the data were done using a newly developed analysis tool, the CRisp SPectral EXplorer (CRISPEX, further details in Appendix~\ref{sec:tools}), that is primarily aimed at imaging spectroscopy data.
The observational data are introduced in the following Sect.~\ref{sec:obsandred}, after which the results of a statistical analysis of these flows are presented in Sect.~\ref{sec:results}. 
Section~\ref{sec:discandconc} offers a discussion of the results obtained in the light of similar and possibly related phenomena such as the inverse Evershed effect and coronal rain, after which conclusions are drawn in Sect.~\ref{sec:conc}.

\section{Observations, data reduction and analysis methods}\label{sec:obsandred}
\subsection{Observational Setup}\label{sec:obssetup}
The \Halpha\ data used in this study were obtained with the CRisp Imaging SpectroPolarimeter 
\citep[CRISP,][]{2008ApJ...689L..69S}, 
at the Swedish 1-m Solar Telescope
\citep[SST,][]{2003SPIE.4853..341S} 
at La Palma (Spain). 
In the setup for these observations the light from the telescope is first guided through an optical chopper and a wavelength selection prefilter before entering the CRISP instrument (a dual Fabry-P\' erot interferometer (FPI) system with a transmission FWHM at the \Halpha\ wavelength of 6.6\,pm and wavelength tuning within a spectral line on the order of $\lesssim$\,50\,ms). 
The chopper ensures synchronization of the exposures obtained and for the observations considered here, the \Halpha\ line was selected by using a prefilter centered on 6563.8\,\AA\ with a FWHM of 4.9\,\AA.
After CRISP, the light beam is split by an orthogonally polarizing beam splitter onto two cameras.
In addition, before entering the FPI but after the prefilter, a few percent of the light is redirected to a third camera, which serves as a wide-band anchor channel in the image post-processing.
All three cameras are high-speed low-noise Sarnov CAM1M100 CCD cameras operating at a frame rate of 35 frames per second with an exposure time of 17\,ms.

\subsection{Data Acquisition and Reduction}\label{sec:data}
\begin{figure*}[hbtp]
  \includegraphics[width=\textwidth]{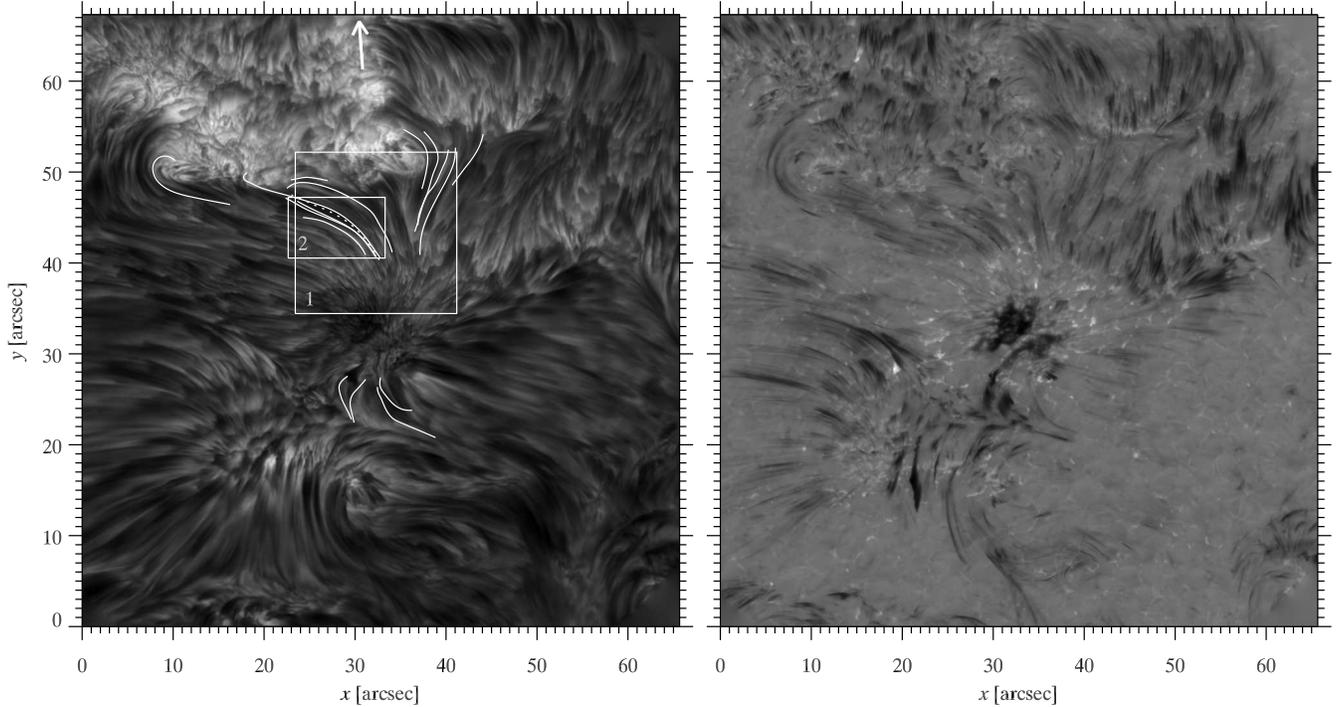}
  \caption{
	Sample images. 
	\emph{Left}: full field-of-view in \Halpha\ line center. 
	The loop paths defined with CRISPEX along which space-time diagrams have been extracted are overlaid.
	The dashed path is discussed in detail in \ref{sec:res_projvel} and Fig.~\ref{fig:timeslice_ID24}.
	The numbered white boxes indicate regions of interest that are discussed further in Sect.~\ref{sec:res_obsprop} (cf.~Figs.~\ref{fig:timeseries_roi1} and \ref{fig:spectprop}, respectively).
	The arrow at the top of the panel indicates the direction of the limb and 	the line-of-sight (pointing away from the observer). 
	\emph{Right}: same field-of-view in \Halpha$-$0.7\,\AA.
	}
    \label{fig:twopanel}
\end{figure*}

The \Halpha\ sequence in question was acquired on 2008 June 11 between 7:55--8:32\,UT. 
The data consist of a time series of profile scans sampling the \Halpha\ line at 23 wavelengths, ranging between $\pm$1.1\,\AA\ around line center at 0.1\,\AA\ spacing. 
At each wavelength a burst of $8$ exposures was taken, thus resulting in an overall cadence of 6.2\,s per profile scan.
The field-of-view covers an area of about 67$\times$67 arcsec$^{2}$ and contains a small sunspot with a rudimentary penumbra (AR~10998), located at $(X,Y)$~$\approx$~($-$684,$-$152), corresponding to $\mu$=0.67.
Figure~\ref{fig:twopanel} shows samples of the observed region in \Halpha\ line center (\emph{left panel}) and the blue wing (\Halpha$-$0.7\,\AA, \emph{right panel}).
The pixel size is 0.071\,arcsec\,px$^{-1}$, well below the SST's Rayleigh diffraction limit for \Halpha, which is 0.17\,arcsec.

With the use of real-time tip-tilt correction and the adaptive optics system at the SST 
\citep{2003SPIE.4853..370S}, 
as well as the image post-processing technique Multi-Object Multi-Frame Blind Deconvolution 
\citep[MOMFBD,][]{2005SoPh..228..191V}, 
the quality of observations can be greatly improved, the latter being employed in order to remove the remaining high-order seeing effects.
To that end, all images (i.e., at each wavelength within the profile scan) were divided into 64$\times$64\,px$^{2}$ overlapping subfields to be processed as a single MOMFBD restoration. 
In that process, the wide-band exposures act as anchors that enable the precise alignment of the restored narrow-band CRISP images. More information on the MOMFBD post-processing of similar data sets can be found in
\citet{2008A&A...489..429V}. 

After MOMFBD restoration, the data was further corrected for the transmission profile of the wide-band prefilter and for the effects of diurnal rotation of the image that result from the alt-azimuth design of the SST. 
Finally, the images are also de-stretched following 
\citet{1994ApJ...430..413S},	
which removes most of the remaining small-scale rubber-sheet seeing effects.

\subsection{Flocculent Flows}\label{sec:res_obsprop}
Closer inspection of the image sequence in either wing of \Halpha\ shows the presence of flows along what, by comparison with the line center image, appears to constitute the chromospheric canopy (including as such the superpenumbral fibrils of the small sunspot). 
The flows are neither uniform nor continuous but rather flocculent, i.e., the flows are observed as dark condensations or blobs flowing along the chromospheric fibril structures.

\begin{figure*}[hbtp]
  \includegraphics[width=\textwidth]{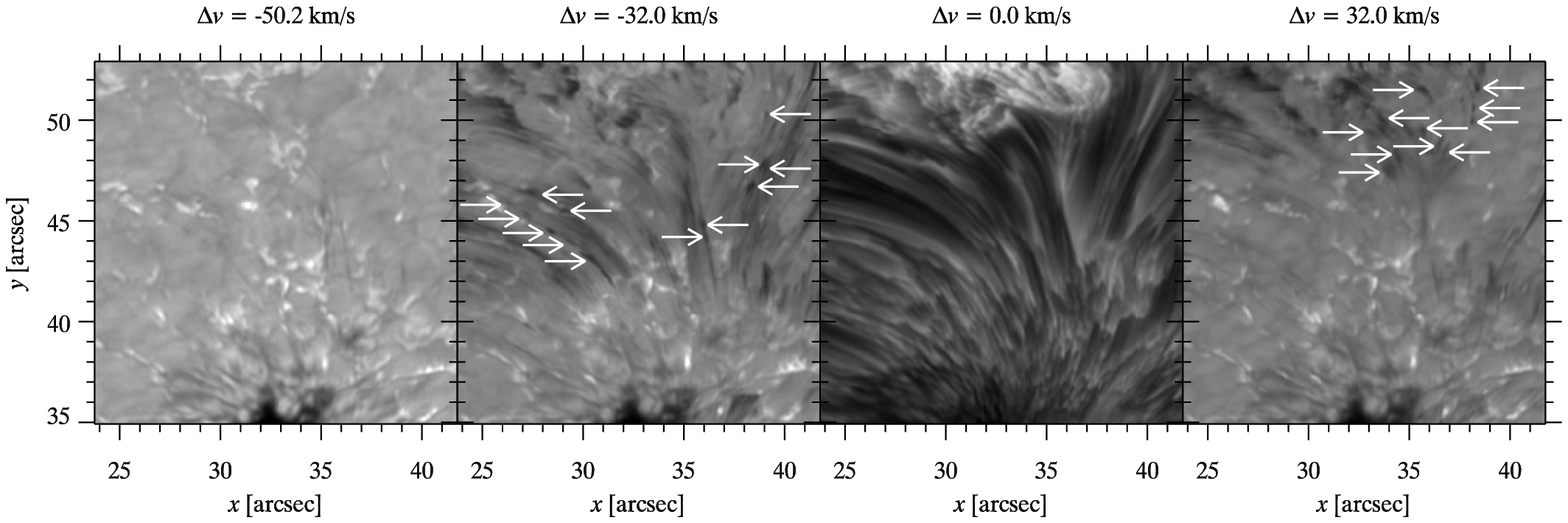}
  \caption{  
  	Cutout from Fig.~\ref{fig:twopanel} of region of interest (ROI) 1 at different wavelengths:
  	\Halpha$-$1.1\,\AA\ ($\Delta v = -50.2$\,\kms, \emph{leftmost panel}), 
  	\Halpha$-$0.7\,\AA\ ($\Delta v = -32.0$\,\kms, \emph{middle left panel}), 
  	\Halpha\ line center itself (\emph{middle right panel}) and
  	\Halpha+0.7\,\AA\ ($\Delta v = +32.0$\,\kms, \emph{rightmost panel}). 
  	The arrows indicate some examples of inflowing and outflowing features.
  	A movie of this figure is provided electronically.}
    \label{fig:timeseries_roi1}
\end{figure*}
Figure \ref{fig:timeseries_roi1} (and in particular the movie of that figure) shows examples of flocculent features (arrows). 
Note that the features in the middle left panel are not visible in the rightmost panel and vice versa, i.e., the same features are not visible in both wings simultaneously.
The apparent coincidence of the trajectories of these flocculent features with superpenumbral fibrils as well as the main flow direction are reminiscent of the properties of the inverse Evershed flow. 
It should be noted, however, that these flocculent flows occur more generally (i.e., also along fibrils that connect different network locations) and not necessarily only propagating towards the sunspot.
Although some of these outward moving features can be found near the sunspot, they typically appear further away from the sunspot, towards a downflow footpoint anchored in the network.
The features occur typically in streams, either throughout the whole observing period or, what appears to be more common, in recurrent, shorter episodes of intermittent appearance. 

Some fibrils remain dark in the wings of \Halpha\ throughout most of the time series or show slower dynamics than the flocculent flows.
The right-hand panel of Fig.~\ref{fig:twopanel} shows a number of these. 
In some cases the flocculent features and these constantly present dark fibril structures seem superimposed on each other, whereas in other cases also fibrils move down towards the sunspot, albeit at smaller velocities (cf.~for instance the middle left panel of the online movie).

\begin{figure}[hbtp]
  \includegraphics[width=\columnwidth]{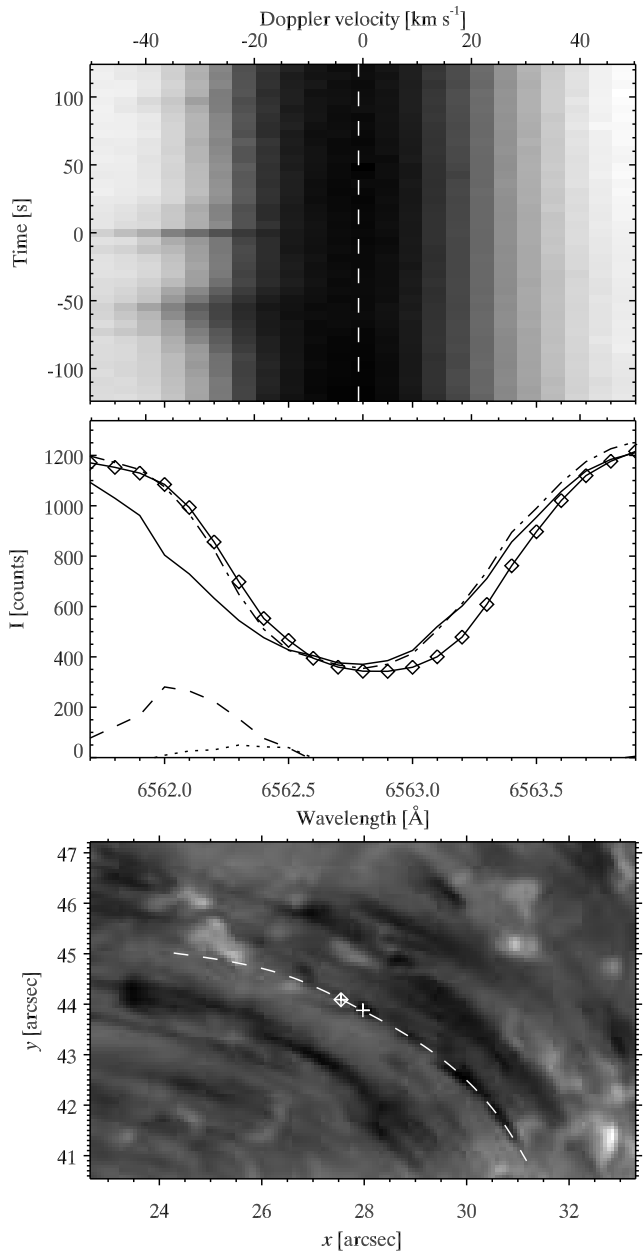}
  \caption{  
	Spectral properties of a typical flocculent feature occurring in ROI~2.
	\emph{Top panel}: spectrum-time diagram at the location marked with a white cross in the bottom panel.  
  	$t=0$ is defined as the instant sampled by the middle and lower panels. 
  	Around $t=-50$\,s another flocculent feature crossed that location.
  	The white dashed line indicates the average line core shift over the displayed time range.
	\emph{Middle panel}: spectral profile (\emph{solid line}) and difference profile (the spectral profile 	subtracted from the field-of-view averaged profile, \emph{dashed line}) of the flocculent feature. 
	A similar set of profiles (\emph{dash-dotted} and \emph{dotted} lines, respectively) is shown for the 		neighboring location, without a flocculent feature, specified by the cross-diamond in the bottom panel. 
	In addition, the the full field-of-view averaged profile (\emph{solid curve with diamonds}) is shown.
 	\emph{Bottom panel}: intensity image at \Halpha\ $-$0.7\,\AA, showing an example of a flocculent feature.
  	The dashed curve specifies the path of the flocculent feature.
 	}
    \label{fig:spectprop}
\end{figure}

Figure \ref{fig:spectprop} shows spectral properties of a typical flocculent feature.
The bottom panel shows the feature in an intensity image in the wing of \Halpha\ at a shift of $-$0.7\,\AA\ (or equivalently, $\Delta v\!=\!-$32.0\,\kms). 
During its visibility in the data sequence, the feature moves along an arched path from the left to the right through the field-of-view, curving down to the lower right of the subfield shown (towards the sunspot which lies just outside this subfield on the lower right side), with similar other features preceding it and features in adjacent fibril structures.
The middle and upper panels of Fig.~\ref{fig:spectprop} illustrate that the flocculent features are most clearly observed in the \Halpha\ wings.
The flocculent flows are composed of features with typical Doppler shifts between 20--40\,\kms\ in either the blue or red wing of \Halpha, depending on their propagation direction with respect to the observer.
Note that, the features appear to show up as a separate component to the local spectrum (the dashed line), as illustrated in the middle panel.
In this particular case, where an inflow towards the sunspot is observed, the separate component appears to be superimposed upon an overall shift of the line core.
Though the extent to which this separate component is visible in the spectrum may vary with location, time and propagation direction with respect to the line-of-sight, it is a typical property of the flocculent flows.

\subsection{Methods}\label{sec:methods}
\subsubsection{Velocities and Accelerations}\label{sec:methods_dynamics}
\begin{figure}[hbtp]
  \includegraphics[width=\columnwidth]{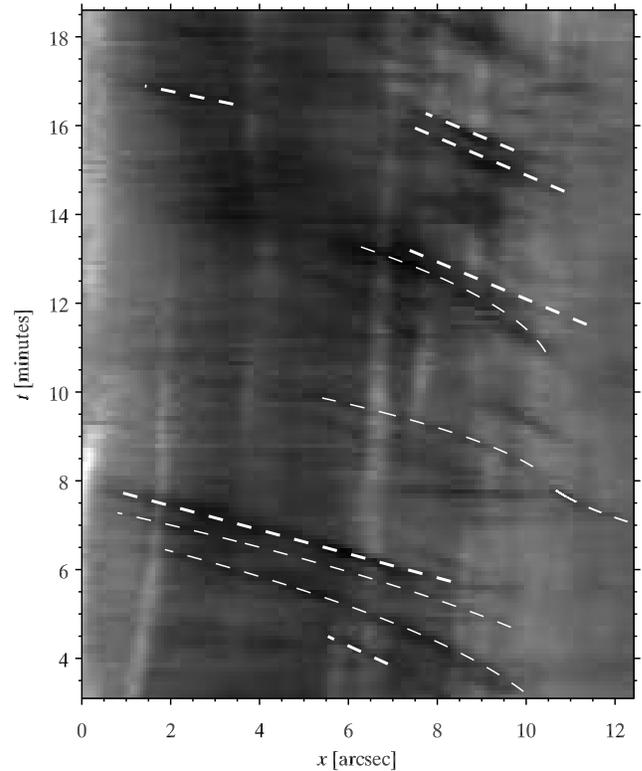}
  \caption{  
  	Part of the space-time diagram extracted along dashed track in Fig.~\ref{fig:twopanel}. 
  	The space-time diagram was extracted at \Halpha$-$0.7\,\AA\ ($\Delta v = -32.0$\,\kms). 
  	The dashed curves indicate measurements of the projected velocity and acceleration components and 			illustrate the method discussed in Sect.~\ref{sec:res_obsprop}.
 	}
    \label{fig:timeslice_ID24}
\end{figure}

Using the path-drawing functionality in CRISPEX, flocculent features were manually traced along their visible trajectories and space-time diagrams were subsequently extracted along these paths.
The left-hand panel of Fig.~\ref{fig:twopanel} shows the final selection of the determined paths overlain on the \Halpha\ line center image, while Fig.~\ref{fig:timeslice_ID24} shows such an extracted space-time diagram corresponding to the dashed track in the former figure.
As the features appear as dark streaks in these diagrams, the determination of their projected velocities and accelerations is a relatively trivial task, that was further simplified by using the CRISPEX auxiliary program TANAT. 

The majority of the dark streaks in the extracted space-time diagrams are straight, and hence a single measurement suffices to obtain the projected velocity component in those cases. 
However, some features have curved space-time tracks; these are fitted with a parabola to three points along them, yielding estimates of the corresponding acceleration or deceleration.
Dashed lines have been overlain in Fig.~\ref{fig:timeslice_ID24} to illustrate this procedure.
A measure for the error in the projected velocities and accelerations was determined by repeatedly measuring randomly-selected dark streaks in each space-time diagram resulting in a measurement uncertainty of 4.0\,\kms\ and 63\,\mss\ for the velocity and acceleration, respectively.

The Doppler velocity of the separate component was obtained employing two methods, a determination of the first moment with respect to wavelength and a Gaussian fit, both to the difference profile (cf.~the middle panel of Fig.~\ref{fig:spectprop}).
In the first we followed 
\citet{2009ApJ...705..272R} 
(see also Eq.~(1) therein),
where the first moment method was applied to Rapid Blue Excursions (RBEs).
The integration is only performed over the range where $I_\lambda^{\rm{avg}} - I_\lambda$ is positive (i.e., the feature is in absorption).
For the Gaussian fit method, the same constraint is enforced and only the spectral points over that same range are used in fitting a single Gaussian with the height, position and width of the peak as the three parameters to be fitted.
For both methods, an estimate of the error in each Doppler velocity determination was obtained by repeating the calculations with boundaries progressively shifted inwards by one wavelength position at a time.
The error for that particular Doppler velocity was then set to the standard deviation of those measurements, resulting in an average error of 0.7\,\kms\ and 4.3\,\kms\ for the first moment and Gaussian fit method, respectively.

While the separate component constitutes in large part the line-of-sight signal of a flocculent feature, it is typically superimposed on an overall shift of the line core, which may be indicative of a lower velocity flow at the same location.
Both a Gaussian and a 4$^{\rm{th}}$ order polynomial were therefore fitted to the core of the line (typically the middle 11 wavelength positions) at each pixel and for each time step. 
Comparison of the resulting fits with the actual local line profile showed the 4$^{\rm{th}}$ order polynomial to be more robust in fitting the core of the \Halpha\ line than the Gaussian. Hence, the results from the polynomial fits were used to construct a Doppler map for each timestep, where the offset between the local minimum in the line profile and that in the average line profile was converted to a Doppler shift velocity. 
This is also a habitual way in which velocities for the inverse Evershed effect have been obtained in spectroscopic studies.

\subsubsection{Morphological Parameters}\label{sec:methods_morphology}
The flocculent features' lengths and widths were determined as follows. 
For each flocculent feature, the intensity profile along the spatial axis was extracted from the space-time diagram at the wavelength closest to its average Doppler velocity and subtracted from the field-of-view averaged intensity at that wavelength.
Next, a difference profile was determined by subtracting the average background intensity profile (i.e., the profile averaged over the time intervals in the space-time diagram where no flocculent features were measured) from the feature's intensity profile.
The difference profile was subsequently lowered by the smallest of the two minima closest to the maximum representing the feature location, as not doing so typically resulted in an overestimation of the FWHM.
Finally, a Gaussian fit to the resulting profile was performed, yielding a FWHM that was converted to a length. 
However, in order to ensure that only flocculent features (rather than the background) were measured, those measurements for which the feature FWHM was smaller than the FWHM resulting from a fit to the background profile (over the same spatial points) were rejected.
This procedure was then repeated for each point along the velocity measurement, yielding a set of length measurements for each flocculent feature, the average of which was taken to be the length of the feature in question.
A similar procedure was adopted to obtain feature widths, but using the intensity image instead, i.e., stepping along the traced path of the feature while it propagates and extracting the intensity profile perpendicular to (rather than along) the propagation direction.

Since the values of the fitted parameters are influenced by the number of points included in the Gaussian fit (i.e., the more points included, the larger the resulting FWHM typically is), both the length and width profile were fitted over three different ranges.
Lengths were calculated by taking 11, 21 and 31 points into account, while the width profile was fitted with 5, 11 and 17 points, in both cases centered on the coordinates where the feature was measured.
The FWHM resulting from the different fits for which the 1-$\sigma$ value is lowest and equal to or smaller than 1\,px was then selected as the feature size, while the standard deviation between the three results was taken to be a measure of the uncertainty in the size determination.
The average error for the lengths is 44\,km, while that for the widths is 30\,km.

\section{Results}\label{sec:results}
\subsection{Dynamics of the Flocculent Flows}\label{sec:res_dynamics}
\subsubsection{Projected Velocities}\label{sec:res_projvel}
\begin{figure}[hbtp]
  \includegraphics[width=\columnwidth]{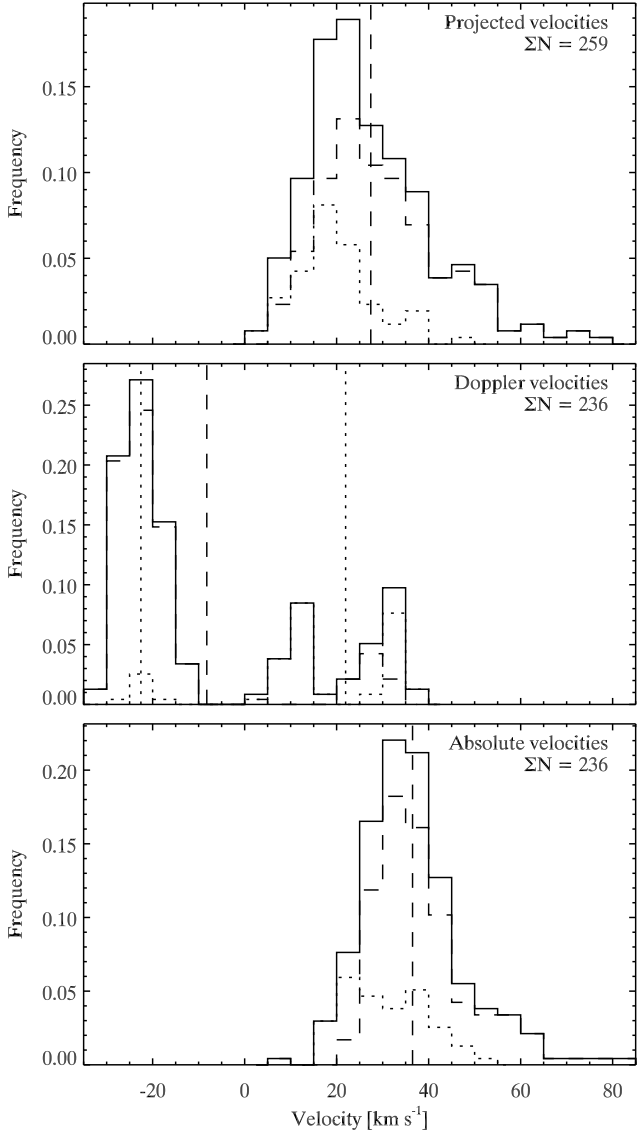}
  \caption{  
  Velocity histograms of the flocculent features with a bin size of 5\,\kms. 
  {\em{Top panel}}: Projected velocities obtained from measurement in the space-time 
  diagrams.
  {\em{Middle panel}}: Doppler velocities. 
  {\em{Bottom panel}}: Absolute velocities obtained by combining the projected and Doppler velocities through vector addition.
  In each panel the solid distribution describes all measurements, the dashed one the subsample of features moving towards the sunspot, the dotted one those moving away.  
  The vertical dashed lines indicate the average velocity all considered measurements, while the dotted lines in the middle panel show the average Doppler velocity of the positive and negative subsamples.
 	}
    \label{fig:velocities}
\end{figure}

In total we obtained 259 velocity measurements of which 188 can be identified with features moving towards the spot, the remaining 71 features moving in the opposite direction.
As pointed out before, the latter features are observed mostly towards the footpoints in the network, rather than closer to the sunspot, however still in structures that appear connected to the sunspot.
Figure~\ref{fig:velocities} shows in the upper panel a frequency histogram of the projected velocity determinations.

The distribution of all measurements peaks around 20\,\kms, with a tail extending up to almost 80\,\kms, resulting in an average velocity of 27.4\,\kms. 
There is a significant difference between inflow (towards the sunspot) and outflow, yielding average velocities of 30.2\,\kms\ and 20.0\,\kms, respectively.
The high-velocity tail of the distribution is dominated by features moving towards the sunspot, while the lower velocities are relatively more abundant for features moving in the opposite direction.

\subsubsection{Doppler Velocities}\label{sec:res_dopvel}
As the Gaussian fit requires a minimum of three data points, this results in fewer successful Doppler velocity determinations than for the first moment method, reducing the total number of measurements by about 18\%, compared to the reduction of about 9\% with the first moment approach.
Nevertheless, both the distributions of Doppler velocities are very similar regardless of the method, their averages falling within each-other's measurement error as determined above.
On one hand, this validates the first moment approach, while showing on the other hand that the influence on the derived statistics of the reduced number of valid measurements for the Gaussian fit method is small.

The middle panel of Fig.~\ref{fig:velocities} shows the overall Doppler distribution, as well as that of the subsamples for features moving towards and away from the sunspot.
The determined Doppler velocities range between $-$31.4\,\kms\ and 36.6\,\kms, with an average of $-$8.2\,\kms\ (or $-$16.5\,\kms\ and 14.6\,\kms\ when considering the inflowing and outflowing subsamples, separately).
The blueshift peak is dominated by features moving towards the sunspot, as expected from the orientation of the majority of the tracks shown in Fig.~\ref{fig:twopanel} with respect to the slanted line-of-sight.
Conversely, features that move away from the sunspot along these tracks are redshifted.
The average Doppler shift for the blueshifted features is $-$22.6\,\kms, with the average of the subsamples differing by less than about 0.3\,\kms\ from the total average.
The redshifted features show a double-peaked distribution, centered around 10\,\kms\ and 35\,\kms, respectively, and with an average of 21.9\,\kms.
As the contribution of the outflowing features is mostly to the lower of those two peaks, its average is correspondingly lower than that of the inflowing features, that contribute almost exclusively to the higher velocity peak.

\subsubsection{Absolute Velocities}\label{sec:res_absvel}
Combining the projected and Doppler component velocities produced the distributions in the bottom panel of Fig.~\ref{fig:velocities}.
Both the shapes of the distributions and range in velocities are similar to those in the upper panel.
The absolute velocities found range between 7.2--82.4\,\kms, with an average of 36.5\,\kms\ when taking all measurements into account.
Considering the inflow and outflow subsamples separately, the averages are 38.8\,\kms\ and 30.3\,\kms, respectively.
Propagating the uncertainties determined for the projected and Doppler velocities to the absolute velocity, the average error in the absolute velocity was found to be 5.9\,\kms. 

\subsubsection{Accelerations}\label{sec:res_acc}
\begin{figure}[hbtp]
  \includegraphics[width=\columnwidth]{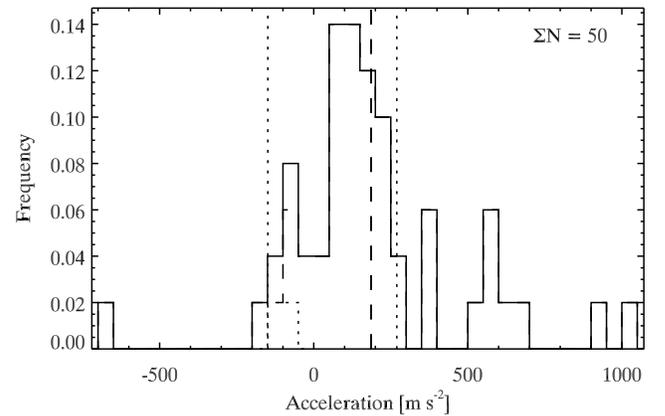}
  \caption{  
  Histogram of the acceleration measurements. 
	\emph{Solid line}: all measurements.
	\emph{Dashed line}: features moving towards the sunspot.
	\emph{Dotted line}: features moving away from the sunspot.
	The vertical lines indicate the average over all measurements (\emph{dashed}), as well as decelerations and accelerations (\emph{dotted}).
	The bin size is 50\,\mss.
 	}
    \label{fig:acchist}
\end{figure}

Figure \ref{fig:acchist} shows the frequency histogram for the 50 acceleration measurements.
Of these, 40 correspond to accelerations, whereas 10 are of decelerations (of which 8 are of features moving towards the sunspot and 2 of features moving in the opposite direction).
The average acceleration (i.e., only including positive values) is 270\,\mss, which lies close to the gravitational acceleration in the solar photosphere of 274\,\mss.
However, this coincidence is fortuitous since most values are much smaller; the average is boosted by the eight outliers with values in excess of 400\,\mss.
Note that the motion of the flocculent features is likely field-guided, so that only the component along the fibrillar path acts as effective gravity.

The average deceleration is much smaller (149\,\mss), with an only slightly higher value of 162\,\mss\ when considering the features moving towards the sunspot and a yet smaller value (96\,\mss) for features moving towards the network.

\subsubsection{Lifetimes and Occurrence Frequency}\label{sec:res_occur}
Lifetimes were determined using the start- and endpoints of each space-time feature track as in Fig.~\ref{fig:timeslice_ID24}. 
\begin{figure*}[hbtp]
  \includegraphics[width=\textwidth]{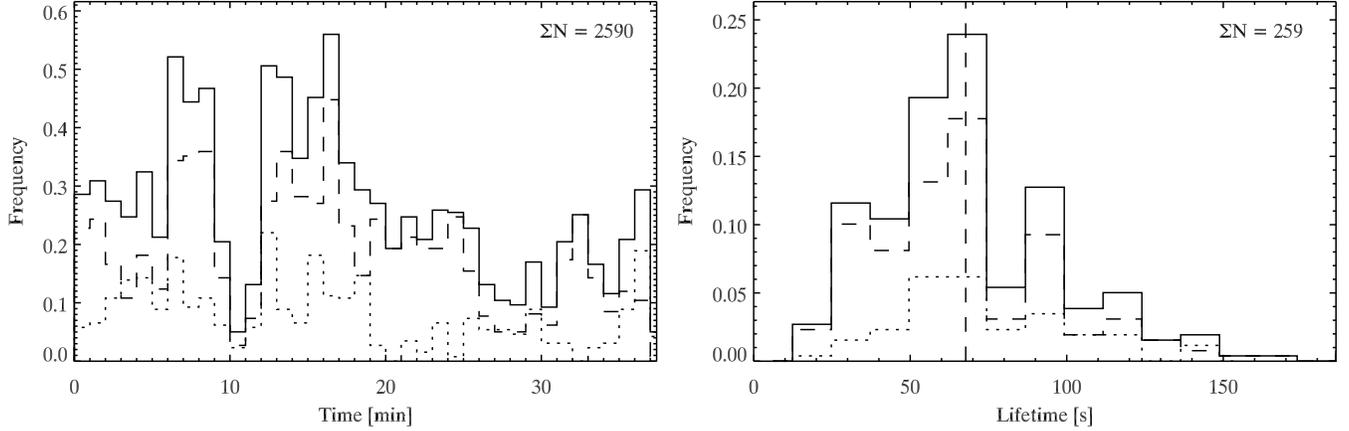}
  \caption{  
  Occurrence frequency (\emph{left panel}, 1-min bins) and lifetimes (\emph{right panel}, 12.4\,s bins).
  \emph{Solid line}: all measurements.
  \emph{Dashed line}: features moving towards the sunspot.
  \emph{Dotted line}: features moving away from the sunspot.
  The vertical dashed line in the right-hand panel indicates the average lifetime. 
 	}
    \label{fig:occur_life}
\end{figure*}
The right-hand panel of Fig.~\ref{fig:occur_life} shows the resulting distribution. 
The 12.4\,s lower cutoff results from requiring at least two samplings for a velocity measurement.
The mean visibility lifetime is 67.7$\pm$8.8\,s, with no significant difference between in- and outward motion.
Note that the lifetimes are rather lower limits, as the features become obscured by fibrils at line center when the Doppler velocity becomes small.

The occurrence frequency was derived by counting the number of measurements at each single time step, resulting in the histograms shown in the left-hand panel of Fig.~\ref{fig:occur_life}.
On average 68 features can be observed each minute, with an average occurrence rate of 47 and 20 features per minute for the in- and outflow subsamples, respectively.
The occurrence frequency appears to decrease somewhat during the second half of the observations, after a peak of 145 occurrences in one minute.

\subsection{Shapes of Flocculent Features}\label{sec:res_sizes}
\begin{figure*}[hbtp]
  \includegraphics[width=\textwidth]{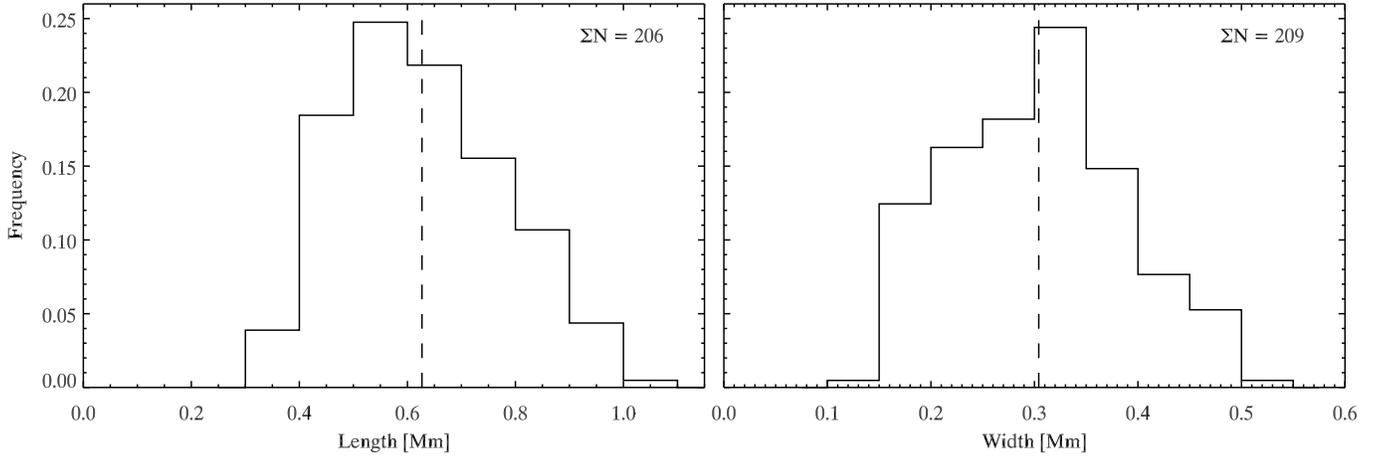}
  \caption{
  Size histograms showing flocculent feature lengths (\emph{left panel}, bin size of 100\,km) and widths 
  (\emph{right panel}, bin size of 50\,km).
  }
    \label{fig:sizes}
\end{figure*}

Figure~\ref{fig:sizes} shows the size distributions of the flocculent features. 
Out of 236 features with successfully determined Doppler velocities, lengths could be determined for 206 features, while 209 widths were obtained.
The lower cutoff in both distributions is at the telescope diffraction limit. 
Upper cutoffs were visually set at 20\,px (or about 1\,Mm) for the lengths and at 10\,px (equivalent to slightly over 500\,km) for the widths.
On average, the flocculent features measure 627\,km in length and 304\,km in width, but lengths range between 353--1005\,km, while widths of 141--509\,km are found.

We also determined eccentricities for those 186 features with both length and width measurements.
In general the features are largest in the propagation direction; only 3 features have a larger ``width'' than ``length''.
On average, the eccentricity of the flocculent features is 0.83$\pm$0.05, with the same value for the average for the larger prolate sample of 183 features, while the average for the oblate sample of 3 features is much smaller (0.27$\pm$0.05).

The morphology of the flocculent features changes only slightly as they progress along their trajectories.
Some features appear to get stretched in length as they move towards the sunspot, while others appear to be compressed.
Slightly more features are found to expand rather than shrink in length as they propagate.

\subsection{Doppler Shift of the Line Core}\label{sec:res_lineshift}
\begin{figure}[hbtp]
  \includegraphics[width=\columnwidth]{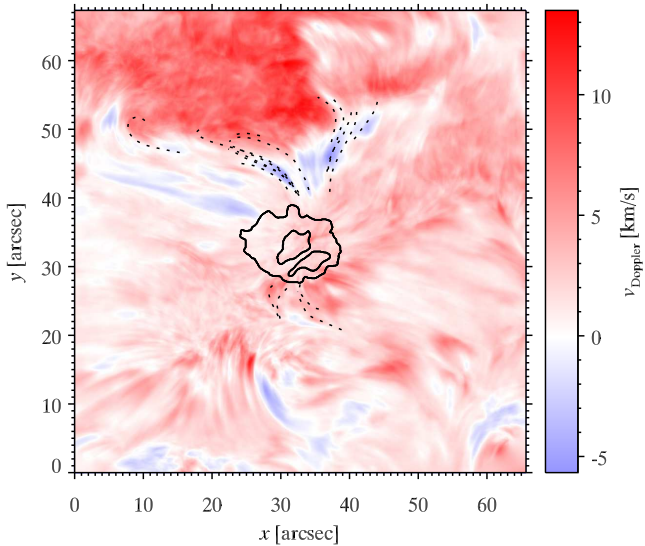}
  \caption{  
  Time-averaged core Doppler shift map of the full field-of-view. 
	\emph{Red}: positive Doppler velocity (redshift).
	\emph{Blue}: negative Doppler velocity (blueshift).
	\emph{Dotted}: the same tracks as in Fig.~\ref{fig:twopanel}.
	\emph{Solid}: contours for the umbra-penumbra boundary and the outer penumbra boundary at \Halpha\ $-$1.1\AA.
 	}
    \label{fig:dmapfig}
\end{figure}

Figure~\ref{fig:dmapfig} shows the time-averaged result of the core Doppler shift measurements. 
The average values lie between about $-$5.7\,\kms\ and 13.5\,\kms, where the minus sign (and blue color) indicates motion towards the observer. 
The average Doppler velocity over the full field-of-view is 2.7\,\kms, directed away from the observer.

There is a conspicuous asymmetry near the sunspot between blueshifted patches on the limb-ward side and red-shifted ones on the disk-center side.
These illustrate general inflow on the order of a few kilometers per second, with slightly higher velocities of up to 10\,\kms\ in the redshifted patches and up to $-$5\,\kms\ in the blueshifted ones.
The blueshifted limb-ward patches generally coincide well with the ends of the trajectories of the flocculent flows, as well as with the superpenumbral fibrils observed in \Halpha\ line center (cf.~Fig.~\ref{fig:twopanel}).
On the disk-center side such correlation is less strong.
The temporal evolution of line core Doppler shift maps shows the blueshifted patches on the limb-ward side of the sunspot to be relatively stable, while the redshifted patches on the disk-center side are more variable, being replaced by blueshifts for several minutes at a time before returning to a redshift.

Space-time diagrams extracted from the Doppler map cube along the same flocculent flow trajectories show only occasionally enhancements that are co-spatial with the dark streaks observed in the space-time diagrams from the regular intensity data.
In those cases for which corresponding enhancements are found, the related (absolute) line core Doppler shifts are typically on the order of 5--15\,\kms.
Also, the sign of the core Doppler shift correlates well with that of the flocculent feature Doppler velocity.

\section{Discussion}\label{sec:discandconc}
Our key finding is the presence of high-speed flows towards and away from the sunspot that are of intermittent, flocculent nature rather than continuous. They are observed in the \Halpha\ wings and appear to follow the fibrillar chromospheric structures seen in the \Halpha\ core.
Summarizing their properties:
\begin{enumerate}
	\item	
	We observe dark flocculent features with an average length of 627$\pm$44\,km and width of 304$\pm$30\,km 	propagating through the chromospheric canopy. 
	
	\item	
	An average absolute velocity of 36.5$\pm$5.9\,\kms\ is found, but velocities in the range of 5--80\,\kms\ 	are also measured. 
	Features moving away from the sunspot propagate at a significantly lower average velocity of 30.3\,\kms, 	against an average inflow at 38.8\,\kms.
	
	\item	
	On average, flocculent features can be observed for 67.7$\pm$8.8\,s and typically 68 features are visible 	each minute. 
	The actual lifetimes may be considerably larger, as the features typically (dis)appear in mid-flight or 	before reaching the photosphere as observed in the far wings.
	
	\item	
	Though practically co-spatial, the flocculent flows appear to be a phenomenon distinct from the 			``classical'', lower-velocity inverse Evershed effect. 
\end{enumerate}

In this section we first discuss these flows with respect to the inverse Evershed effect (Sect.~\ref{sec:disc_invevershed}), coronal rain (Sect.~\ref{sec:disc_rain}), and then discuss the possible driving mechanism (Sect.~\ref{sec:disc_driving}).

\subsection{Flocculent Flows and Inverse Evershed Effect}\label{sec:disc_invevershed}
The general picture of inflow in the chromospheric superpenumbra is confirmed by the results obtained in this study.
Not only is it observed as an inflow at line-of-sight velocities of a few kilometers per second (cf.~Sect.~\ref{sec:res_lineshift}), but also in the form of flocculent features moving at considerably higher velocities. 
Although we find slightly higher average values of up to 15\,\kms, the line core Doppler shift velocities are similar to the inverse Evershed effect velocities obtained in earlier studies.
Already
\citet{1913ApJ....37..322S}	
found inverse Evershed velocities on the order of about 3\,\kms\ on average and similar velocities were obtained by
\citet{1978SoPh...57...65B},	
who measured an average radial line-of-sight velocity of 1.8\,\kms\ from $\rm{Mg\,b}_{I}$ spectra.
\citet{1990A&A...233..207D}	
studied \Halpha$\pm0.3$\,\AA\ velocity maps and found velocities of 2.6\,\kms, while
\citet{1988A&A...201..339A}	
obtained chromospheric line-of-sight velocities (averaged over the position angles considered) between +2\,\kms\ (i.e., upflow) and $-$4\,\kms, and
\citet{1969SoPh....9...88H}	
found an average velocity vector peaking at the penumbral rim at 6.8\,\kms.
The velocities measured for the flocculent features, are more on the order of the chromospheric inflow velocities 
\citet{1975SoPh...43...91M}	
and
\citet{1962AuJPh..15..327B}	
found in their filtergram studies (velocities of 5\,--\,15\,\kms\ and 40\,--\,50\,\kms, respectively), but are also comparable to the velocities of some of the single absorbing elements measured in the spectroscopic study by 
\citet{1969SoPh....9...88H},	
who found velocities ranging between $-$35 and +50\,\kms\ (where positive values indicated downflows).

We suggest nevertheless that most of the older studies, hampered by much lower spatial resolution than reached here and considering their morphological description of the inflows, concern the more gentle flow patterns that we observe as patches of enhanced line core Doppler shift seen in Fig.~\ref{fig:dmapfig}.
The flocculent features described here are in fact not observable at line center where they seem to be obscured by overlying fibrils (cf.~Fig.~\ref{fig:timeseries_roi1}).
The flow channels of the inverse Evershed effect and the flocculent flows appear to be distinct and are thus not as such the same flows.

\subsection{Flocculent Flows and Coronal Rain}\label{sec:disc_rain}
Morphologically, the flocculent flows are reminiscent of coronal rain. 
Recent work by 
\citet{2012ApJ...745..152A}	
and in particular
\citet{2012_Antolin_etal}	
shows this is not only the case qualitatively, but also quantitatively.
Both studies consider similar \Halpha\ imaging spectroscopy data obtained with CRISP as for this study.
\citet{2012ApJ...745..152A}	
suggest that the coronal rain phenomenon might be a common one, with single condensations attaining average absolute velocities on the order of 70\,\kms\ and typical sizes of 0.74\,Mm in length and 0.31\,Mm in width.
Similar dynamics and morphological quantities have been found by 
\citet{2012_Antolin_etal}	
for on-disk cases of coronal rain, including also in the currently considered data set.
As these similarities may lead to difficulties in distinguishing both phenomena, as well as being indicative of a possibly related driving mechanism, we discuss this phenomenon in more depth.

Typical velocities for the coronal phenomenon are on the order of a few tens to a few hundred kilometers per second, as for instance found by
\citet{2001SoPh..198..325S} 
(studying quiescent coronal loops) and
\citet{2007A&A...475L..25O}, 
who found Doppler shifts of 50\,--\,100\,\kms\ and velocities of at least up to 100\,\kms\ from \HeI\ and \OIV\ data obtained with the Coronal Diagnostics Spectrometer on SOHO.
\citet{2005A&A...436.1067M} 
and
\citet{2005A&A...443..319D} 
obtained velocities in a similar range (30\,--\,120\,\kms) from observations, although the simulations of 
\citet{2005A&A...436.1067M} 
yielded typical velocities of about 50\,\kms, while also indicating that subsequently downflowing blobs may travel at increasingly higher velocities of up to 230\,\kms\ due to evacuation of the loop by preceding condensations.

The flocculent features observed in this study are somewhat smaller than those found by
\citet{2012ApJ...745..152A}	
and
\citet{2012_Antolin_etal},	
however mostly so in length, and move at much lower average velocities than coronal rain.
The lower velocities may result from the flocculent flow channels being much shorter than coronal loops, as the traced trajectories range between 3--14\,Mm.
\citet{2003A&A...411..605M}	
reported on simulations of 10\,Mm coronal loops, but found propagation velocities on the order of only a few kilometers per second for the condensations that formed.

A further, important, difference is that flocculent features appear to constitute a footpoint-to-footpoint, rather than apex-to-footpoint, flow.
Even though no flocculent feature could be traced all the way from one footpoint to the other in this data set, the (re)appearance of the features close to the fibril apices moving at considerable velocities would argue against them forming through catastrophic cooling alone.
In that respect, the higher flocculent flow velocities compared to the velocities from 
\citet{2003A&A...411..605M}	
can be understood, as coronal rain accelerates from zero velocity as it starts precipitating. 

Yet, even though the exact mechanism causing coronal rain may not be applicable to the flocculent flow phenomenon, a similar morphology may be nonetheless indicative of a thermal non-equilibrium process leading to the observed condensations.
Comparison with the results obtained by
\citet{2012ApJ...745..152A}	
shows yet another striking similarity between both phenomena: in a majority of the flocculent flow cases adjacent fibrils display blobs propagating in the same direction at the same time, which may point at co-evolution of the conditions in neighboring fibrils.

\subsection{Driving Mechanism}\label{sec:disc_driving}
Distinguishing wave phenomena from actual mass flows is a recurring challenge, as what is observed as flocculent features may actually be gas compressed by propagating wave fronts, their velocities reflecting the phase speed of the waves, rather than actual mass motion.
Indeed, similar features that have been observed before by
\citet{2006ApJ...648L..67V},	
who studied \Halpha\ line center data (i.e., no further spectral information), were explained as MHD waves.
The qualitative morphology and behavior of the ``moving features in loops'' is practically identical to the flocculent flows described here and the differences with the projected velocities they report (40--120\,\kms) may well be attributed to projection effects.
With typical sizes of up to 275\,km, the features they analyzed are notably smaller, however. 
Also, they report on bright blobs at line center, whereas the flocculent flows can only be seen in the wings of \Halpha, despite the viewing angles being comparable for both data sets.
This may, however, be due to an inherent difference in fibril topology or by a different orientation of those fibrils to the line-of-sight. 
A more recent study by
\citet{2011_Lin_etal},	
covering in part the same data sets, expands on the earlier results by
\citet{2006ApJ...648L..67V}	
by including CRISP data.
They show that the bright blob-like features typically have small Doppler shifts ($\pm$5--10\,\kms), last for up to about 150\,s, and slightly increase in length as they propagate.

Features with similar morphology and behavior have also been found in other studies, albeit propagating at lower velocities. 
For instance,
\citet{2008A&A...486..577S}	
found chromospheric magneto-acoustic waves in \Halpha\ observations. 
These waves manifest themselves mainly as (trains of) blobs with average dimensions as small as 725\,km (1\,arcsec) in length and 360\,km (0.5\,arcsec) in width, propagating at phase velocities between 12--42\,\kms.
Both the sizes and the phenomenology are similar to what we observe for the flocculent flows.
However, in contrast to those (but in common with some of the features described by
\citet{2006ApJ...648L..67V}),	
the blobs observed by
\citeauthor{2008A&A...486..577S}	
show displacement perpendicular to their propagation direction. 
In a more recent study of \CaII~8542\,\AA\ data obtained with {\em{SST}}/CRISP
\citet{2010MmSAI..81..693W}	
reported on (among other things) bright blobs propagating at projected velocities of 4--8\,\kms\ along neighboring fibrils that he explained as a wave manifestation.

Considering this scenario further, we note that typical chromospheric sound speeds are on the order of 10--15\,\kms\
\citep[e.g.,][]{1988A&A...201..339A,	
2000A&A...357..735T,	
2008A&A...486..577S,	
2011_Lin_etal},	
while Alfv\' en speeds may range between 175--600\,\kms\ for magnetic field strengths between 30--100\,G
\citep[cf.][and references therein]{2008A&A...486..577S},	
and a reasonable chromospheric density of 2$\times10^{-13}$\,\gcm\
\citep[cf.][and references therein]{2000A&A...357..735T}.	
Although subsonic flocculent features have been found, most exceed the chromospheric sound speed by far and hence the flocculent flows cannot be attributed exclusively to acoustic waves.
A magneto-acoustic wave scenario may allow for much higher velocities (even comparable to flocculent flow velocities), as for instance shown by 
\citet{1975SoPh...44..299G},	
who measured phase velocities on the order of 70\,\kms\ in \Halpha\ observations of quiet chromosphere and a sunspot.
Assuming typical chromospheric number densities and a magnetic field strength of 10\,G, these velocities could be reconciled with an Alfv\' en wave scenario. 
However, one thus needs to assume rather low field strengths (i.e., below 6\,G) in order to obtain an Alfv\' en velocity as low as the average flocculent flow velocity of 36\,\kms.
In addition, the high Doppler velocities of the flocculent features in combination with the changes in the Doppler component velocities as the features propagate along their curved paths, would seem to argue for actual mass motion, rather than waves. 

An alternative driving mechanism could be found in the siphon flow model. 
Siphon flows have been suggested before in the context of the inverse Evershed effect
\citep[cf.][]{
1969SoPh....9...88H,	
1975SoPh...43...91M,	
1988A&A...201..339A,	
1990A&A...233..207D,	
1992A&A...259..307B,	
1993SoPh..145..257K,	
1997Natur.390..485M}.	
In this model, a flow is driven by a gas pressure inequality between the footpoints of a magnetic flux tube, which in turn is caused by a difference in magnetic field concentration (and hence strength) at the respective footpoints.
A higher magnetic field strength at one footpoint will result in a lower gas pressure at that same footpoint and vice versa.
As the magnetic field strength of a sunspot will be larger than in the surrounding plage, a flow directed towards the sunspot would naturally occur in this model.
Analytical solutions predict both sub- and supersonic flows, depending on the parameters of the loop structure, however supersonic flows in the critical solution (i.e., where the flow has been accelerated from sub- to supersonic) will always return to a subsonic state through the formation of a shock
\citep[cf.][]{
1980SoPh...65..251C, 
1988ApJ...333..407T}.	
Since the flocculent flows appear to be footpoint-to-footpoint flows, the siphon flow model may offer a viable way to explain them.  
For most cases a supersonic flow would be required, 
at the very least for that part of the fibrils where the flocculent features are visible.

A number of problems arise, however. 
The siphon flow model may explain the general characteristic of an inward flow towards the sunspot, but it does not explain why flocculent features could be formed.
Although shocks will occur in the transition from supersonic to subsonic flows and have been observed to a larger or smaller degree of confidence
\citep[cf.][]{%
1993A&A...279L..29D, 
2006ApJ...645..776U, 
2012A&A...537A.130B},	
there is no evidence that these shocks cause propagating condensations and would rather show up as brightenings due to material compression and the consequent rise in temperature.
A few cases of decelerating flocculent features were found in this study, but the return to subcritical velocities through a shock was not evident from the data.
However, as 
\citet{2012A&A...537A.130B}	
already pointed out for their case, these shocks could be occurring at lower altitudes that are not sampled by the \Halpha\ line.
Alternatively, the shocks could be obscured from view by overlying fibrils, in a similar way to flocculent features closer to \Halpha\ line center.
Also, we observe flows towards the plage, i.e., in the direction opposite to what the siphon flow model predicts.
Although in most cases these features are seen close to the plage (footpoints) only and could thus be propagating in fibrils that are actually not connected to the sunspot (allowing possibly low enough field strength for oppositely directed flows), this does not hold true for all cases.
Nonetheless, as suggested by
\citet{1980SoPh...65..251C}, 
an oppositely directed flow may occur in a fibril with an established siphon flow as a result of a change in pressure difference. 
Evidence for the existence of such a counterflow has for instance been provided by
\citet{2011ApJ...743L...9G},	
in a recent study of a pore and its surroundings.

Although considering much larger scales, recent simulations described by
\citet{2011A&A...532A.112Z}	
may also offer a possible driving mechanism. 
The simulations they present show the emergence of a blob-like feature from the transition region, resulting from the build-up of pressure (in turn caused by a heating event at one footpoint) propelling the feature further through the corona along a magnetic loop structure before falling down again.
The feature described has much larger dimensions than typical flocculent features or coronal rain and only occurred once during a one hour simulation run. 
A similar heating mechanism might nevertheless work on smaller scales too, producing blobs of smaller sizes propagating through the chromosphere and/or transition region at possibly shorter time intervals.

\section{Concluding remarks}\label{sec:conc}
In this paper we have analyzed flocculent flows occurring along the chromospheric canopy as observed at high spatial, temporal and spectral resolution in \Halpha\ data obtained with the CRISP instrument at the Swedish 1-m Solar Telescope.
Although we have been able to obtain statistical properties of a considerable sample of flocculent features, we can at this point merely speculate about the mechanism driving the flocculent flows.
The results obtained, both qualitatively and quantitatively, seem to argue against a purely wave-related phenomenon.
Rather, the flocculent features are more likely to occur as part of an already present (supersonic) siphon flow, which could develop its flocculence as a result of, for instance, recurring heating events leading to the formation of condensations (in analogy to the larger scale phenomenon described by
\citet{2011A&A...532A.112Z}),	
or waves interacting with the flow-carrying fibrils, causing the condensations to form as the fibril is pushed out of thermal equilibrium.
In either case, these condensations would subsequently be carried by the siphon flow at supersonic velocities (where they are observed) before being decelerating through a shock to subsonic velocities upon reaching the photosphere.

Given that this phenomenon appears to be ubiquitous and not solely related to the superpenumbra, a study of multiple data sets (including also more quiet regions) and with different diagnostics would not be out of place.
In addition, where observations may constrain the properties of the flocculent features as well as the necessary environmental conditions for them to occur, simulations would be very helpful in further uncovering their driving mechanism.

\acknowledgements
The authors would like to thank Patrick Antolin for many useful discussions and are indebted to Rob Rutten for numerous helpful suggestions regarding the manuscript.
G.V. would in addition like to thank Luc Rouppe van der Voort, Patrick Antolin, Bart de Pontieu, Sven Wedemeyer-B\" ohm, Mats Carlsson, Jorrit Leenaarts, Eamon Scullion and Jaime de la Cruz Rodriguez for their invaluable input to improve and extend CRISPEX.
G.V. has been supported by a Marie Curie 							
Early Stage Research Training Fellowship of the European CommunityÕs 6th Framework Programme (MEST-CT-2005-020395): The USO-SP International School for Solar Physics. 
The Swedish 1-m Solar Telescope is operated on the island of La Palma by the Institute for Solar Physics of the Royal Swedish Academy of Sciences in the Spanish Observatorio del Roque de los Muchachos of the Instituto de Astrof{\'\i}sica de Canarias.

\appendix
\section{Analysis tools: introducing CRISPEX and TANAT}\label{sec:tools}
Sophisticated instruments such as the CRISP instrument at the SST deliver large data sets that are often formatted as multidimensional data cubes.
These are nontrivial to browse and analyze, and require therefore a tool with which one may quickly explore the data obtained for interesting features and which, preferably at the same time, allows for more in-depth analysis of the data set of interest.
A widget based tool, programmed in the Interactive Data Language (IDL), was developed with that goal in mind and driven by the particular challenges faced as a result of the extensive observational possibilities offered by CRISP. 
The CRisp SPectral EXplorer (CRISPEX)\footnote{Both the actual distribution and more detailed descriptions of the functionality and options of CRISPEX and its auxiliary routines can be found at http://bit.ly/crispex.}
can currently handle:
\begin{itemize}
	\item	single spectral scans;
	\item	simple 3D temporal cubes (i.e., a time series of intensity images);
	\item	spectrotemporal cubes (i.e., cubes with both spectral and temporal information); and
	\item	Stokes data cubes (i.e., spectrotemporal cubes including also multiple Stokes parameters).
\end{itemize}
As the typical size of CRISP data cubes exceeds the available random access memory on regular laptop or desktop computers, the data are read in to an associated variable and hence CRISPEX need neither read nor keep a full data cube in memory, thereby enabling quick browsing in the spatial, temporal and spectral domain.
This swiftness comes at a cost, however.
In order to have full spectral browsing functionality, a spectral data cube (which is nothing more than a reordered image data cube) must be supplied as well, thus requiring double the amount of disk space for one data set.

Though originally designed and intended to handle CRISP data, CRISPEX can cope with any (observational or synthetic) data, provided it has been formatted in a certain way.
Indeed, the tool has already been successfully used in several studies based on CRISP data 
\citep{2009ApJ...705..272R,
2009A&A...507L...9W,	
2010ApJ...716..154A,	
2011ApJ...736...71W,	
2011_Lin_etal},	
as well as {\em{Hinode}}/SOT data
\citep{2011ApJ...736..121A}.	

\subsection{Visualization and Analysis Options}\label{sec:tools_browse_analysis}
Depending on the input data, different visualization and browsing options are available to the user.
One can simultaneously:
\begin{itemize}
	\item	play forward and backward in time; 
	\item	step through a spectral profile;
	\item	switch between Stokes parameters;
	\item	blink between image frames, stepping both in temporal and spectral dimension.
\end{itemize}
\begin{figure*}[hbtp]
  \includegraphics[width=\textwidth]{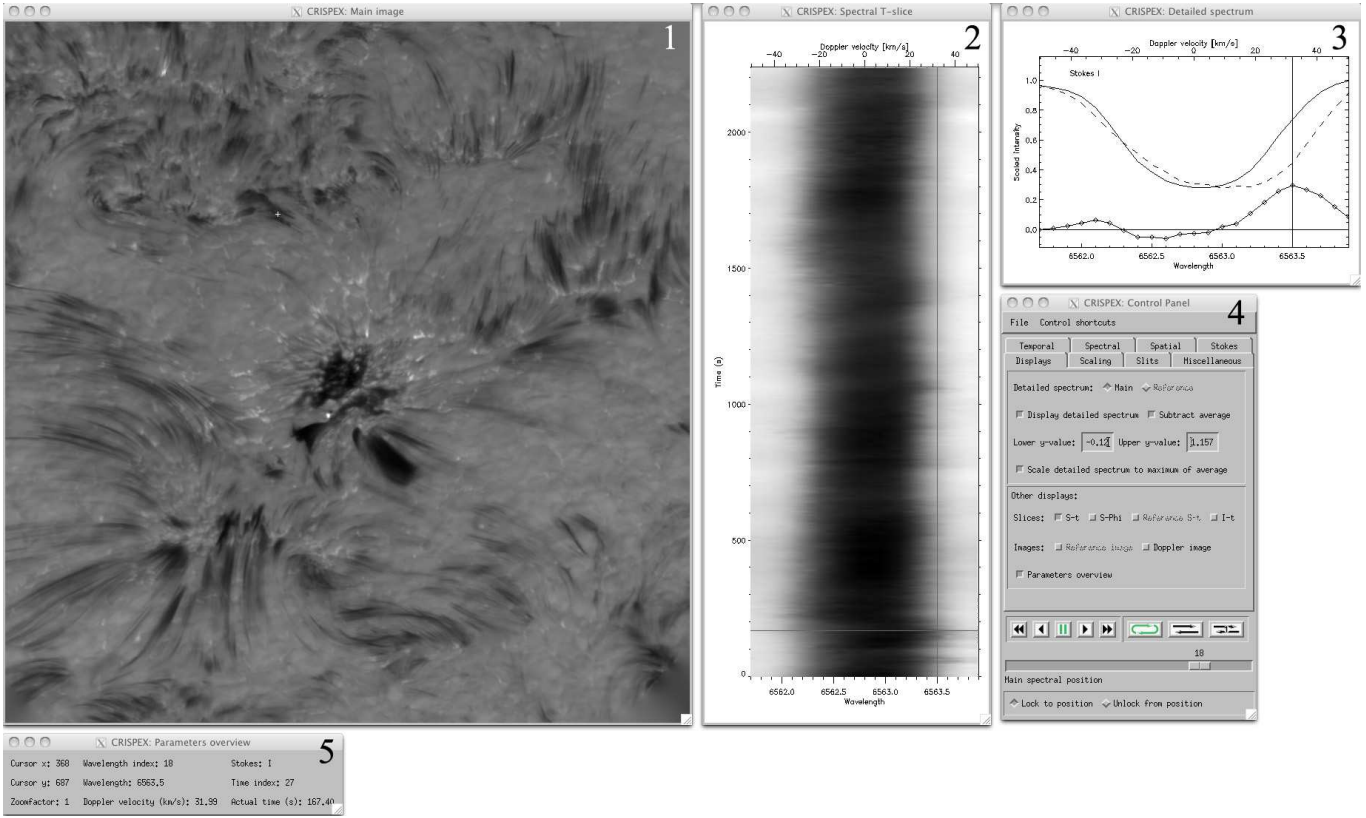}
  \caption{  
  Default layout of windows loaded by CRISPEX, showing clockwise from the top left: (1) the intensity image,
  (2) the spectrum-time diagram, (3) the local spectrum, (4) the control panel and (5) the parameters overview window.
 	}
    \label{fig:crispex_wins}
\end{figure*}
Data feedback is given to the user through multiple screens, of which Fig.~\ref{fig:crispex_wins} shows an example. 
Following CRISP's capability to run multiple line programs (i.e., observing programs where time series in more than one spectral line are obtained), CRISPEX can display two data sets at a time. 
Such reference data (e.g., a reference magnetogram) need not have the same spectral dimension, but are currently only constrained to having the same spatial dimensions and either consist of a single frame or the same amount of temporal frames as the main data cube.
Additionally, CRISPEX allows for simple image and spectrum saving (with the option to include or exclude overlays, such as the cursor or drawn paths), which can be used for simple movie production.

CRISPEX also offers a number of analysis options, but one of its first and most developed features is the extraction of space-time diagrams along user-defined (curved) paths.
These paths can (as they are in this study), but need not necessarily be, manually defined.
One may also supply a file with detections to extract space-time diagrams along, as was done by
\citet{2009ApJ...705..272R}
in their spectral analysis of on-disk counterparts of type II spicules.
Space-time diagrams can be extracted and saved for a single wavelength position or for a given spectral range (which may be the entire spectral scan), as well as for a restricted temporal range.
Another analysis option is given by the spectral slit functionality, which extracts the spectrum for each pixel along a line centered on the cursor.
The slit can be adjusted both in length and in rotation angle, thereby allowing the coverage of practically any feature of interest.
In particular for those cases where a profile scan is available, the in-program Doppler image functionality may be of use to browse the Doppler image constructed from the subtraction of the red from the blue wing.
Though this feature may slow down the overall playback, depending on the memory available on the machine used, one escapes the need to produce a separate Doppler cube.
Lastly, the functionality to display intensity-time diagrams may be particularly useful when cubes do not consist of profile scans but rather of different spectral diagnostics.
This functionality could be used to show, for example, subsequent detection of a phenomenon in different diagnostics of multiple {\em{SDO}}/AIA channels.

\subsection{Auxiliary Programs}\label{sec:tools_aux}
Although the CRISPEX distribution comes with several auxiliary routines, one program in particular deserves some further introduction, since it offers important analysis options complementary to those in CRISPEX.
The Timeslice ANAlysis Tool (TANAT) facilitates the analysis of space-time diagrams extracted with CRISPEX, as it enables the determination of projected velocities and accelerations of features from these diagrams.
Such measurements can subsequently be saved to a file, with the option to add a (custom) flag for each measurement as it is saved, and adding by default a unique identifying number which enables the correct overlay of saved measurements on the corresponding space-time diagram.
In addition, different visualization options allow switching between or combination of space-time diagrams for different positions within the spectral line observed, thereby enabling -- for instance -- the comparison of a feature's signal (or lack thereof) in the red and blue wing of a spectral line.

\bibliographystyle{apj} 
\bibliography{flocflows}

\end{document}